# Differential chiral sensing at exceptional points in intrinsically chiral metastructures

Suzit Hasan Nayem[1], Zhanni Wu[1], Mariusz Drong[1], Kamyar Behrouzi[2], and Boubacar Kanté [1,3,*]

[1]*Department of Electrical Engineering and Computer Science, University of California Berkeley, Berkeley, California 94720, USA*
[2]*Department of Mechanical Engineering, University of California Berkeley, Berkeley, California 94720, USA*
[3]*Materials Sciences Division, Lawrence Berkeley National Laboratory, Berkeley, California 94720, USA*
**Contact author:* bkante@berkeley.edu

**ABSTRACT**. Discriminating opposite enantiomers through their weak chiroptical responses remains a fundamental challenge in optical sensing. Exceptional points (EPs), non-Hermitian singularities at which eigenvalues and eigenvectors coalesce, offer a route to enhanced spectral responses through their characteristic square-root eigenvalue splitting. However, their potential for chiral discrimination remains largely unexplored. We demonstrate exceptional points in an intrinsically chiral hybrid metamaterial and reveal strongly asymmetric responses to enantiomeric perturbations. The interplay between intrinsic chirality and non-Hermitian degeneracy is realized through the coupling of a chiral plasmonic nanoantenna array to a photonic crystal slab. We observe characteristic complex-square-root scaling, with both resonance frequencies and loss-rates exhibiting pronounced handedness dependence. Remarkably, opposite enantiomers produce complex mode splittings that differ by more than an order of magnitude. Our findings establish intrinsic chirality as a new degree of freedom in non-Hermitian physics and reveal a fundamental route toward topological singularities enhanced discrimination of chiral matter.
.

## I. INTRODUCTION

Ever since the discovery of chirality in 1848, it has played a crucial role in modern biochemistry and the evolution of life [1]. The detection and separation of racemic mixtures of enantiomers, *i.e.,* molecules with opposite left and right-handedness, is a challenge in chemistry. The molecular chirality discrimination can be realized with circular dichroism (CD) spectroscopy describing the difference in molecular absorption of left and right-handed circularly polarized light [2]. In general, the molecular CD signal is weak and chiral measurements using such spectroscopic techniques have usually been restricted to relatively high concentrations [3]. It was shown that the enantioselectivity of optical excitation of a molecule is highly dependent on the chirality of the optical field [3,4]. The so-called superchiral near-fields possess high local chiral density and enhanced chiral light-matter interactions. Since then, various efforts have been made in designing artificial nanophotonic structures supporting superchiral near-fields [5-7]. Recently, chiral plasmonic nano-resonators have been proposed as an alternative method towards chiral sensing [8-9]. In this approach, chiral plasmonic nanoparticles support simultaneous electric and magnetic dipole moments. The intrinsic optical chirality of such structures leads to an enantioselective frequency shift when perturbed by chiral analytes. The frequency shift $|\Delta\omega|$ provides information about the chiral perturbation $\delta\chi$ and the sign of the shift (blue or red shift) provides information on the handedness of the perturbation. However, quantifying small concentrations of chiral perturbations has remained challenging.

Exceptional points (EPs) occur when at least two eigenmodes of an open system coalesce [10]. Recent demonstrations have shown that operating a sensor around EP singularities increases the responsivity to external perturbations [5,11-16]. A target nanoscale object that enters the evanescent field of the photonic structure perturbs the system from its exceptional point, leading to enhanced complex frequency splitting. A natural question to ask is whether exceptional points can be implemented to address the challenge of enantiomeric detection at small analyte concentrations. Recent studies have demonstrated refractive index and chiral sensing in achiral media operating at exceptional points (EPs) [5,13,15]. Chiral sensing has also been reported in intrinsically achiral/chiral systems without EPs [9,17]. However, the role of EPs in intrinsically chiral nanostructures for differential enantiomeric detection has remained unexplored, mostly because of the challenge of implementing EPs in a three-dimensional metastructure with intrinsic chirality [18-22]. In this letter, we present the first evidence of the existence of EPs in a metastructure with intrinsic chirality. We show that EPs in such systems enable enhanced differential chiral sensing. Moreover, we derive a closed-form expression linking the complex frequency shift to chiral perturbations of the intrinsically chiral metastructure, establishing a new approach for sensitive enantiomeric detection.

## II. RESULTS & DISCUSSION

The proposed hybrid structure consists of an intrinsically chiral plasmonic array coupled to a photonic crystal slab. The schematic of the hybrid (metallo-dielectric) and chiral structure is presented in Fig. 1. A periodic array of helix plasmonic nano-resonators surrounded by a cladding medium of refractive index, $n_{clad} = 1.3$ is placed on top of a dielectric photonic crystal as shown in Fig 1(a). The 3D unit-cell and its 2D top-view (x-y plane) are shown in Fig. 1(b) and Fig. 1(c), respectively. The helix resonator is made of gold, the permittivity of which is described by a Drude model [23]. The number of turns of the designed helix resonator is chosen to be $n_{turn} = 0.9$ so that the chiral helix turns into a planar non-intrinsically chiral split ring resonator when its height is zero [Fig. 1(d)]. The height of the helix ($h_{helix}$) can be tuned to control the strength of the intrinsic chirality of the plasmonic nano-resonators. When $h_{helix}$ reduces to zero, the helix loses its intrinsic chirality. Figures 1(e-f) present right-handed (Fig. 1e) and left-handed (Fig. 1f) plasmonic helices, respectively. The silicon photonic crystal is a periodic array of circular holes on a glass substrate. The photonic crystal itself is designed to support a fundamental transverse-electric (TE) mode as shown in Fig. S1 of the Supplemental Material [24], which is symmetry-protected at the center of the Brillouin zone. The plasmonic chiral nanoantenna array couples magnetically to the non-radiative photonic mode. The near-field coupling between the plasmonic and photonic modes can be tuned by controlling the in-plane shift between the plasmonic and photonic lattices. The in-plane shift, $dx$, is defined in Fig. 1(c) by the relative positions of the centers of the plasmonic and photonic unit-cells.

We investigate the impact of a chiral perturbation on the plasmonic helix array alone (without the dielectric photonic crystal) using non-Hermitian perturbation theory [26]. The complex frequency shift $\Delta\tilde{\omega}$ under small chiral perturbation, $\delta\chi$ is given by the following expression (detailed derivation in Section II of Supplemental Material [24]):

$$\frac{\Delta\tilde{\omega}}{\tilde{\omega}_0} = \frac{1}{c}\frac{-i\delta\chi\iiint_{V_p}(E_0\cdot H_0+E_0\cdot H_0)d^3r}{\iiint_{\Omega}\left(\frac{\partial(\omega\epsilon)}{\partial\omega}E_0\cdot E_0-\mu H_0\cdot H_0\right)d^3r}+\frac{1}{c}\frac{\delta\chi^2\iiint_{V_p}\left(\frac{1}{\eta}(E_0^n)^2-\eta(H_0^n)^2\right)d^3r}{\iiint_{\Omega}(\frac{\partial(\omega\epsilon)}{\partial\omega}E_0\cdot E_0-\mu H_0\cdot H_0)d^3r} \quad (1)$$

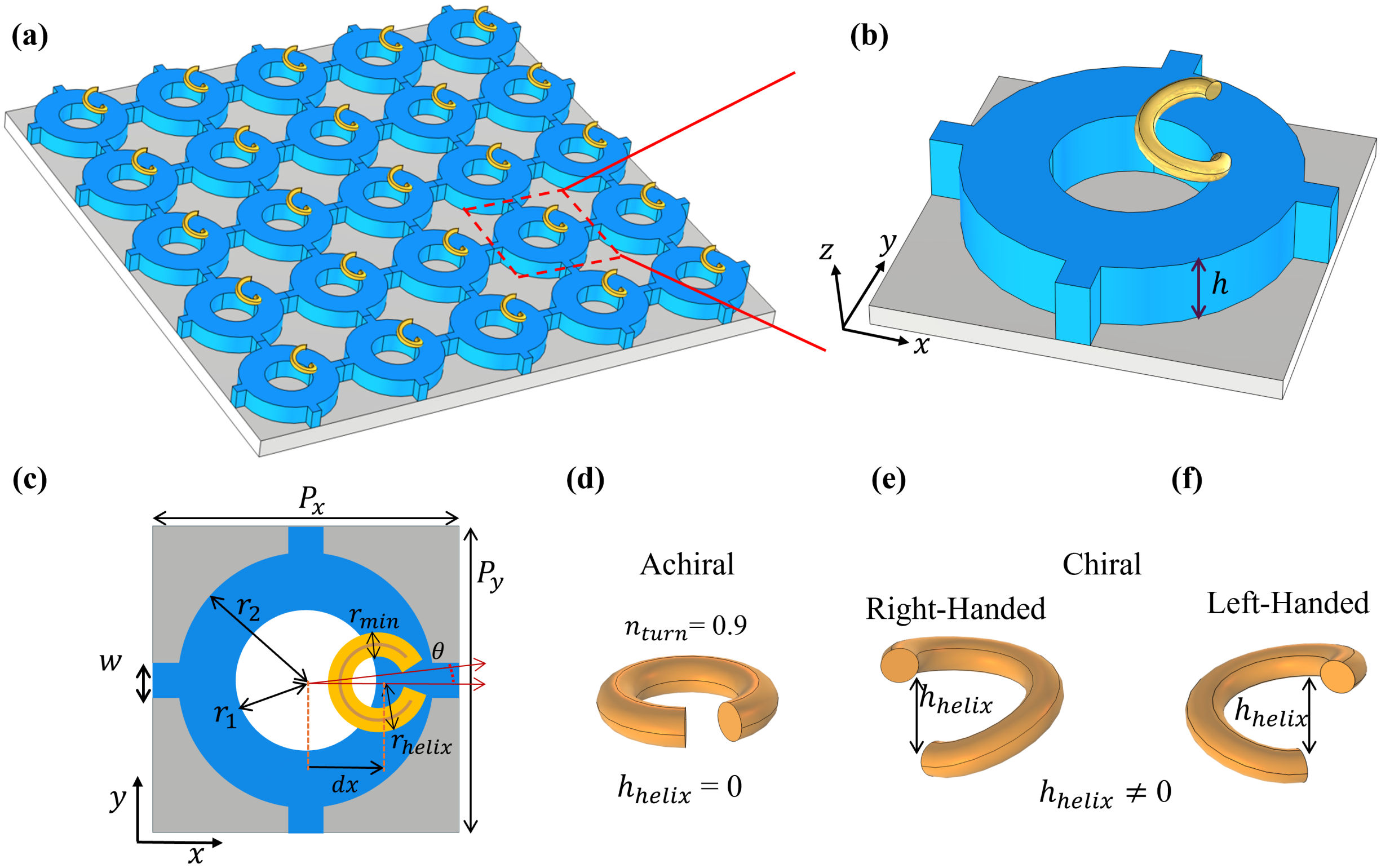

Fig. 1. Hybrid chiral metamaterial (HCM) system architecture (a) Schematic of the hybrid chiral metamaterial (HCM) system. (b) Unit-cells of the HCM made of silicon ($n_{si} = 3.5$) of thickness $h$ on a glass substrate ( $n_{SiO_2}$= 1.5) and a metallic helix surrounded by a cladding medium ($n_{clad} = 1.3$). (c) Two-dimensional (x-y plane) view of the unit-cell with geometric parameters such as the periods $P_x = P_y = 1200\ nm$, the radii $r_1 = 240\ nm$, $r_2 = 500\ nm$, and the width $w = 100\ nm$. The plasmonic helix has a major radius $r_{helix} = 110\ nm$ and a height $h_{helix}$, forming an $360° \times (1 - n_{turn}) = 36°$ in-plane gap where $n_{turn}$ is the number of rotations of the helix, and the in-plane rotation is $\theta = 18°$ with the x-axis. $r_{min} = 30\ nm$. (d) Achiral helix with $h_{helix}$=0 (e) Right-handed chiral helix with $h_{helix} \neq 0$ (f) Left-handed chiral helix with $h_{helix} \neq 0$.

where $(E_0^n, H_0^n)$ are perpendicular components of the unperturbed helix mode. From Eq. (1), it appears that for an achiral resonator under small chiral perturbation, the resonant frequency of the helix nanoantenna array is an even function and it shifts symmetrically with respect to the handedness of the perturbation. In addition, a quadratic frequency shift is expected with $\delta\chi$. This is illustrated in Fig. 2(a) and Fig. 2(b), when $h_{helix}$ is zero (blue curves) and the helix resonator is intrinsically achiral. The frequency shift of the achiral split-ring resonator shows a quadratic frequency shift for small chiral perturbations. When $h_{helix}$ gradually increases, the intrinsic chirality of the helix resonator increases. Figure 2(a) and Fig. 2(b) show that when the intrinsic chirality of the resonator increases, the minimum of the frequency shift moves away from the $\delta\chi = 0$. Equation (1) can be rearranged as: $\frac{\Delta\tilde{\omega}}{\tilde{\omega}_0} = A\,[(\delta\chi - \chi_0)^2 - \chi_0^2]$, where $A = \frac{1}{c}\frac{\iiint_{V_p}\left(\frac{1}{\eta}(E_0^n)^2 - \eta(H_0^n)^2\right)d^3r}{\iiint_{\Omega}\left(\frac{\partial(\omega\epsilon)}{\partial\omega}E_0\cdot E_0 - \mu H_0\cdot H_0\right)d^3r}$. Here, we interpret $\chi_0$, the shift of the location of the dip along $\delta\chi$ axis, as the intrinsic chirality of the structure:

$$\chi_0 = \frac{i\iiint_{V_p}(E_0 \cdot H_0 + E_0 \cdot H_0)d^3r}{2\iiint_{V_p}\left(\frac{1}{\eta}(E_0^n)^2 - \eta(H_0^n)^2\right)d^3r}. \qquad (2)$$

It is straightforward to see that when $\chi_0 \neq 0$, the intrinsic chirality of the resonator leads to an asymmetric frequency shift that depends on the handedness of the chiral perturbation around $\delta\chi = 0$. When $|\chi_0| \gg 0$, *i.e.,* when the resonator exhibits large intrinsic chirality, the frequency shift for a small region around $\delta\chi = 0$ is approximately linear as shown in Fig. 2(a) and Fig. 2(b) for $h_{helix} = 150$ nm. In addition, changing the sign of the handedness of the helix resonator, while keeping all parameters the same, results in a mirror-flipped behavior of the frequency perturbation, as shown in Fig. 2(a) and Fig. 2(b). Ideally, for chiral sensing, a large slope of the frequency shift around $\delta\chi = 0$ is desirable. This can be achieved by increasing the value of $|\chi_0|$, which often results in complex nano-resonator structures. The structural geometric chirality can be quantified using the Continuous Chirality Measure (CCM) framework [27] presented in Fig. 2(c). The CCM is zero for $h_{helix} = 0$ nm and it gradually increases to a maximum for $h_{helix} \approx 410$ nm and subsequently decreases.

We then investigate the hybrid chiral metamaterial (HCM). The system exploits its multi-mode nature, where the plasmonic resonator array couples to the photonic crystal slab (Fig. 1). For an isolated photonic mode which is achiral, the frequency shift under chiral perturbation remains small compared with the response of the plasmonic helix array alone (see Fig. S1 of the Supplemental Material [24]).

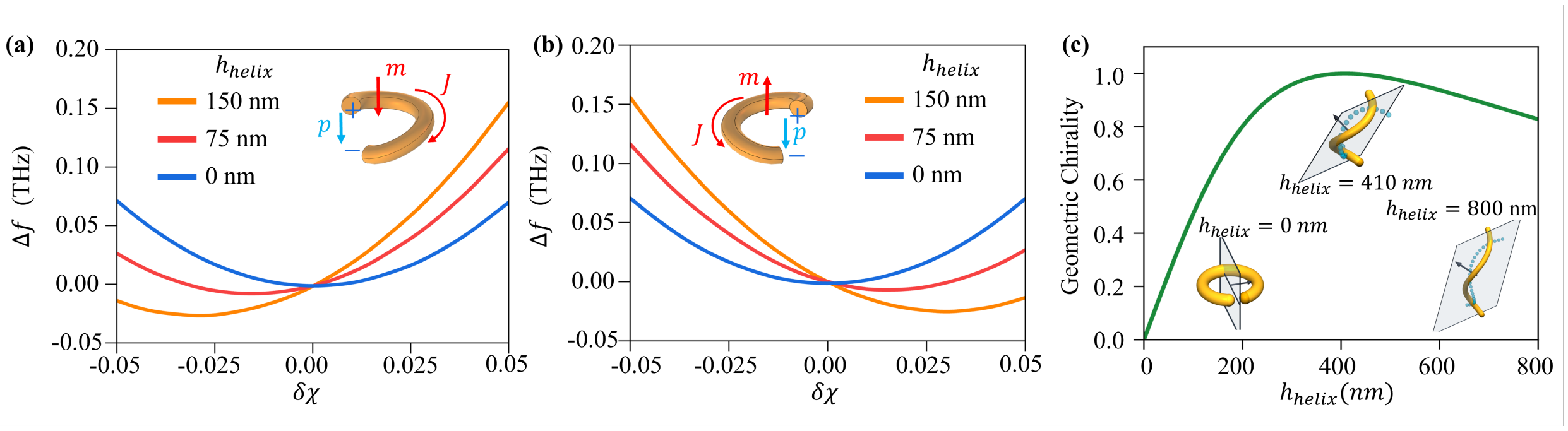

Fig. 2. Chiral perturbation response of the plasmonic crystal. Simulated resonant frequency shift ($\Delta f$) of right-handed helices (a) and left-handed helices (b) plasmonic crystal under chiral perturbation for different values of $h_{helix}$. The inset shows the current flow $J$, the magnetic dipole moment $\boldsymbol{m}$, and the electric dipole moment $p$. (c) Structural Chirality as Continuous Chirality Measure (CCM) for different $h_{helix}$ of HCM structure, with the inset showing the helix, its mirror plane and its mirror image (dotted line) for $h_{helix} = 0\ nm$; $h_{helix} = 410\ nm$; and $h_{helix} = 800\ nm$.

Moreover, the perturbation response is handedness independent. In the HCM, magnetic coupling between the plasmonic and photonic modes dominates due to the quasi-collinear magnetic dipole moments of the PhC-slab and the helix particles. By tuning the thickness ($h$) of the PhC and the relative lateral displacement (dx) between the PhC and plasmonic arrays, the coupling is altered and the system can be driven to an EP. For instance, as seen in Figs. 3(a) and 3(b) for $h_{helix} = 100$ nm, an EP is obtained for $h \approx 345\ nm$ and $dx \approx 187\ nm$. Around the EP singularity, a small chiral perturbation induces a more pronounced frequency splitting compared to the response of the plasmonic array alone. Interestingly, a complementary and large loss-rate splitting is also observed.

To quantitatively understand the response of the HCM, we analytically express the frequency shift around its EP as a function of the chiral perturbation using the resonant state expansion (RSE) method [28]. The resonant states, *i.e.*, the quasinormal modes (QNMs) of the electromagnetic system, are complex eigensolutions of Maxwell's equations satisfying outgoing boundary conditions. The RSE is a rigorous perturbation technique that describes the coupling of an analyte to the HCM in terms of its isolated QNMs. We begin with Maxwell's equations:

$$\nabla \times E = i\widetilde{\omega} B, \qquad (3a)$$
$$\nabla \times H = -i\widetilde{\omega} D, \qquad (3b)$$

where $E$ and $H$ are respectively the electric and magnetic fields, and $\widetilde{\omega}$ is the complex angular frequency. Assuming negligible spatial dispersion in the medium, the electric displacement $D$ and magnetic induction $B$ are given by:

$$D = (\epsilon + \Delta\epsilon)E - i(\chi' + \Delta\chi')H, \qquad (4a)$$
$$B = (\mu + \Delta\mu)H + i(\chi' + \Delta\chi')E, \qquad (4b)$$

where $\epsilon, \mu$ and $\chi'$ are respectively, the unperturbed permittivity, permeability, and normalized chirality index, the latter defined as $\chi' = \chi\sqrt{\epsilon\mu}$ and $\Delta\epsilon, \Delta\mu$ and $\Delta\chi'$ are their arbitrary perturbations, which appear when photonic and plasmonic systems become coupled. Note that $\Delta\chi'$ is directly linked to the intrinsic chirality of the plasmonic helix $|\chi_0|$ (when $|\chi_0| = 0$, then also $\Delta\chi' = 0$) and represents the system's chirality contributing to the PhC-helix coupling. We introduce a complete set of normalized resonant states (RSs) with the complex angular frequencies $\widetilde{\omega}_n$ and fields $E_n$ and $H_n$, satisfying Eqs. (3) and (4) with $\Delta\epsilon = \Delta\mu = \Delta\chi' = 0$, representing isolated resonances of photonic and plasmonic sub-systems [27]. The fields can be expanded using RSs as:

$$E = \sum_n c_n E_n \ , \ \ H = \sum_n c_n H_n \ , \qquad (5)$$

which, after plugging into Eqs. (3) and (4), and performing the projection onto $\{E_n, H_n\}$ the basis, effectively maps Eqs. (3) and (4) onto the $2 \times 2$ matrix generalized eigenvalue problem for two-level systems [17, 28]:

$$\widetilde{\omega}\begin{pmatrix} 1 + V_{11} & V_{12} \\ V_{21} & 1 + V_{22} \end{pmatrix}\begin{pmatrix} c_1 \\ c_2 \end{pmatrix} = \begin{pmatrix} \widetilde{\omega}_1 & 0 \\ 0 & \widetilde{\omega}_2 \end{pmatrix}\begin{pmatrix} c_1 \\ c_2 \end{pmatrix} \qquad (6)$$

*i.e.,*

$$MC = \frac{1}{\widetilde{\omega}}C, \qquad (7)$$

where $C = (c_1, c_2)^T$ and the coupling matrix $M$, which can be thought of as the inverse of the Hamiltonian $H = M^{-1}$, is defined as:

$$M = \begin{pmatrix} \widetilde{\omega}_1 & 0 \\ 0 & \widetilde{\omega}_2 \end{pmatrix}^{-1}\begin{pmatrix} 1 + V_{11} & V_{12} \\ V_{21} & 1 + V_{22} \end{pmatrix} \qquad (8)$$

The matrix elements are:

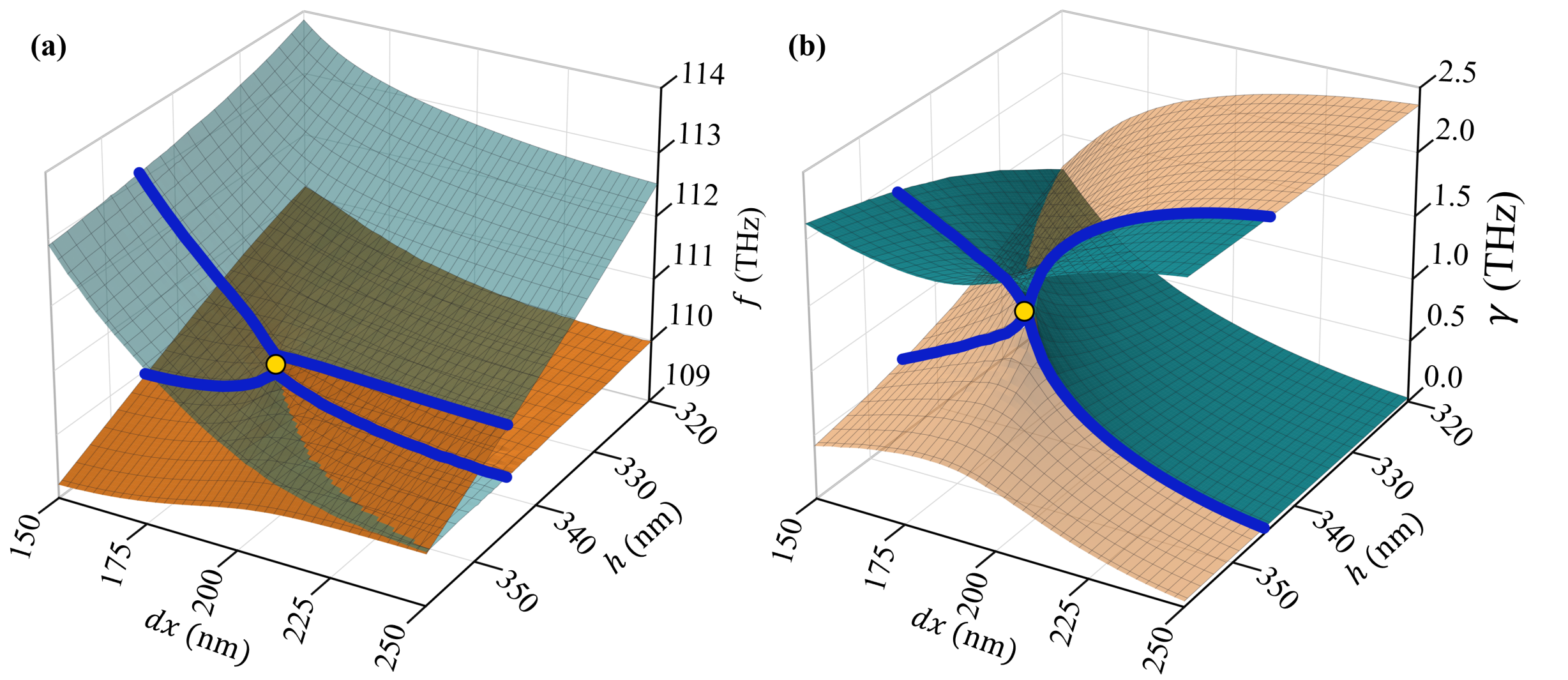

Fig. 3. Evidence of the existence of an EP in the hybrid chiral metamaterial. Real part of the complex frequency [$f$] (a) and imaginary part of the complex frequency [$\gamma$] (b) as a function of $h$ (thickness of the photonic crystal) and $dx$ (shift of the helix from the center of the dielectric photonic crystal) for a right-handed helix with $h_{helix} = 100\,nm$. The blue line is the variation of $f$ and $\gamma$ of the two coupled modes at $h = h_{EP}$ and the yellow dot is the location of the exceptional point

$$V_{nm} = \int [E_n \cdot \Delta\epsilon E_m - H_n \cdot \Delta\mu H_m] d^3r \; - i \int [E_n \cdot \Delta\chi' H_m + E_m \cdot \Delta\chi' H_n] d^3r \quad (9)$$

It can be shown that the exceptional point (EP) *i.e,* $\tilde{\omega}_+ = \tilde{\omega}_- = \tilde{\omega}_{EP}$ occurs when the following equality holds:

$$\left(\frac{1+V_{11}}{\tilde{\omega}_1} - \frac{1+V_{22}}{\tilde{\omega}_2}\right)^2 + 4\frac{V_{12}}{\tilde{\omega}_1}\frac{V_{21}}{\tilde{\omega}_2} = 0 \qquad (10)$$

At this point, assuming that the device is tuned to the EP, we introduce an additional small chiral perturbation $\delta\chi'$, originating, for example, from sensed chiral analytes. Consequently, the matrix elements are perturbed as $V^p_{nm} = V_{nm} + \delta V_{nm}$, where

$$\delta V_{nm} = -i\delta\chi' \int [E_n \cdot H_m + E_m \cdot H_n] d^3r \qquad (11)$$

We derive the splitting of complex frequencies near EP due to chiral perturbation $\delta\chi'$ (see details in the Sec. III of Supplemental Material [24]):

$$\Delta\tilde{\omega} = \tilde{\omega}_+ - \tilde{\omega}_- = 2\pi(\Delta f + i\Delta\gamma) \; = \tilde{\omega}^2_{EP}\epsilon\mu\sqrt{(\alpha + \beta\,\Delta\chi)\,\delta\chi} \qquad (12)$$

The constants $\alpha$ and $\beta$ are determined by the volume integral of eigenmode overlaps between electric and magnetic fields (see details in Sec. III of Supplemental Material [24]). Equation (12) reveals the square-root topology of both frequency ($f$) and loss-rate ($\gamma$) splitting in response to a chiral perturbation. In the absence of intrinsic chirality, the splitting remains symmetric about $\delta\chi = 0$, making the system incapable of enantiomeric discrimination, in full agreement with recent studies of chiral sensing with achiral two-level systems [17]. In contrast, increased intrinsic chirality breaks this symmetry and enhances handedness-resolved responses to perturbations (see

Fig. S2 of the Supplemental Material [24]). To the best of our knowledge, this is the first theoretical analysis showing enhanced handedness-resolved responses to chiral perturbation in intrinsically chiral systems operating near EPs.

To validate this theory, we numerically simulated both resonance frequency splitting ($|\Delta f|$) and loss-rate splitting ($|\Delta\gamma$)) as functions of the chiral perturbation parameter $\delta\chi$ for both right- and left-handed helices based HCM. Increasing the helix height enhances structural chirality up to a critical limit beyond which the helix approaches a straight rod and its chirality declines. We define positive and negative chirality indices for right- and left-handed structures, respectively. As shown in Fig. 4(a), for a right-handed HCM, right-handed perturbations lead to enhanced frequency splitting and suppressed splitting for left-handed perturbations. Conversely, Fig. 4(c) shows the opposite behavior for loss-rates of the same metastructure. The loss-rate splitting under left-handed perturbations is enhanced for the right-handed structure, while it is suppressed under right-handed perturbations.

As the helix height increases from 0 nm to 475 nm, the contrast of both $\Delta f$ and $\Delta\gamma$, *i.e,* ($|\Delta f_{\delta\chi=+}| > |\Delta f_{\delta\chi=-}|$ and $|\Delta\gamma_{\delta\chi=+}| < |\Delta\gamma_{\delta\chi=-}|$) increases monotonically from 0 to 410 nm and reaches peak at 410 nm. This behavior is consistent with the relation between $\Delta\widetilde{\omega}$ and $\Delta\chi$ derived in Eq. (12) and the continuous chirality measure (CCM) shown in Fig. 2(c). The enhancement of discrimination decreases at $h_{helix} > 410\ nm$. For $h_{helix} > 475$ nm, the system fails to reach an exceptional point due to a reduction in mode coupling, caused by a larger portion of the helix volume extending beyond the PhC's evanescent field. The simulation results directly indicate that the enhancement of differential chiral response originates from the increased intrinsic structural chirality.

The same trend is observed for the left-handed HCM depicted in Figs. 4(b,d). Frequency/loss-rate splittings are enhanced for left/right-handed perturbations and suppressed for right/left-handed perturbations. In general, when the handedness of the structure and perturbation matches, the frequency splitting is enhanced, when they mismatch, the loss-rate splitting is enhanced. Thus, by jointly analyzing the frequency and loss-rate responses, the handedness of the perturbing enantiomer can be determined. The additional information provided by the loss channel adds a new degree of freedom for chiral sensing. The ratio between the enhanced and suppressed complex mode splitting magnitude for the same magnitude of chiral perturbation strengths $|\delta\chi|$ is defined as the discrimination factor $D$ as given by the following equation:

$$D_{|\delta\chi|} = \frac{||\Delta\widetilde{\omega}_{enhanced}||}{||\Delta\widetilde{\omega}_{suppressed}||} = \frac{\sqrt{(max(|\Delta f|_{\delta\chi+,}|\Delta f|_{\delta\chi-}))^2+(max(|\Delta\gamma|_{\delta\chi+,}|\Delta\gamma|_{\delta\chi-}))^2}}{\sqrt{(min(|\Delta f|_{\delta\chi+,}|\Delta f|_{\delta\chi-}))^2+(min(|\Delta\gamma|_{\delta\chi+,}|\Delta\gamma|_{\delta\chi-}))^2}} \qquad (13)$$

Discrimination factors, $D$ for right-handed and left-handed HCM structures for different helix heights are shown in Fig.4(e) and (f), respectively. As $h_{helix}$ increases, the structural chirality $\Delta\chi$ increases and leads to higher discrimination ratio, with a maximum around $D \approx 20$ for $h_{helix} = 410\ nm$. Note that the discrimination factor $D$ remains almost constant for large perturbation strength and reaches the same values for both right-handed and left-handed HCM structures for identical $h_{helix}$.

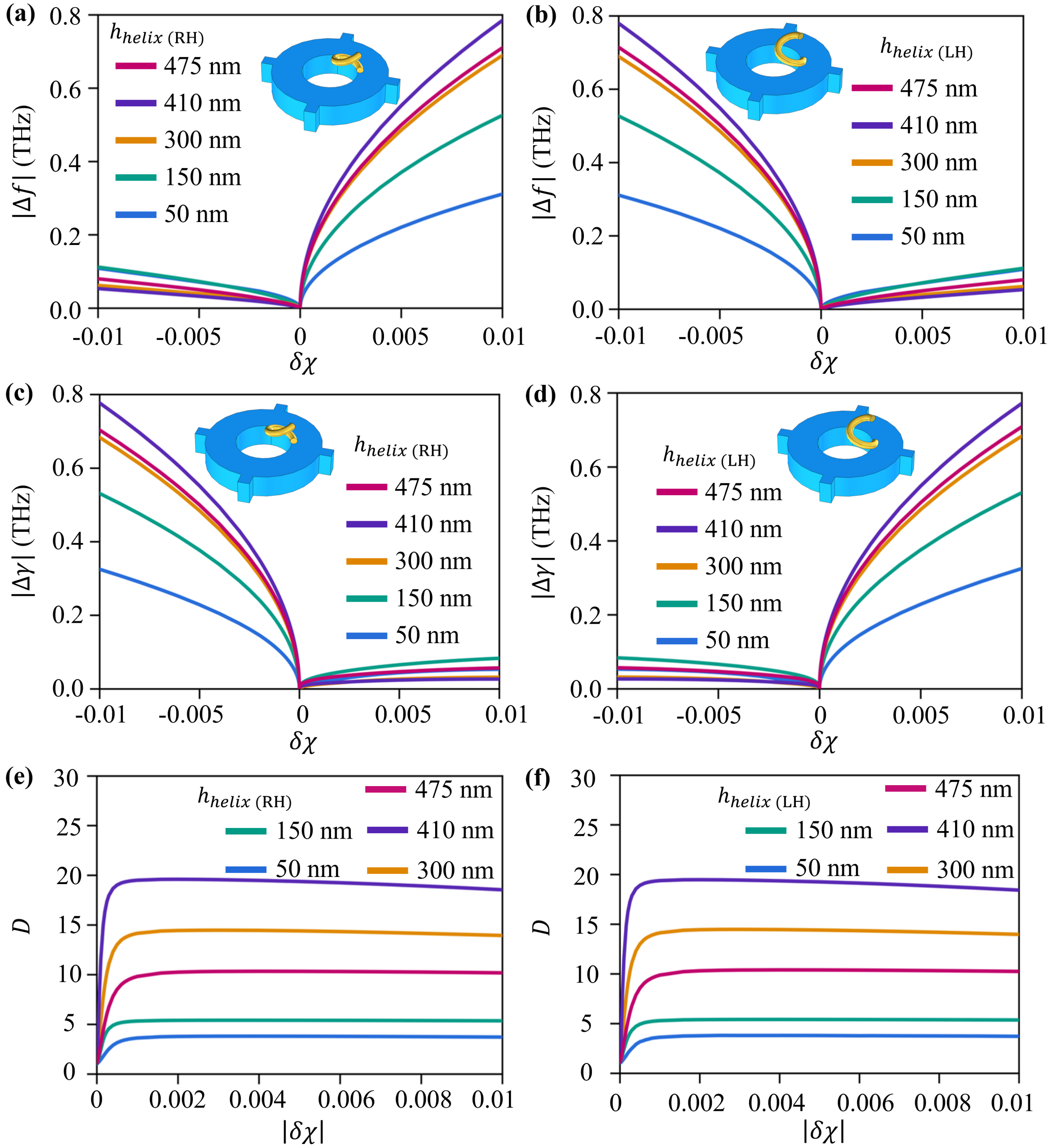

Fig. 4. Complex resonance splitting under chiral perturbation around the EP singularity of hybrid chiral metamaterials (HCM) of different helix heights. Frequency splitting (a), loss-rate splitting (c) and discrimination factor (e) for right-handed helices as a function of the chiral perturbation. Frequency splitting (b), loss-rate splitting (d) and discrimination factor (f) for left-handed helices as a function of the chiral perturbation.

Interestingly, the maximum discrimination factor occurs around the height of the metallic helix with maximum geometric chirality, confirming that intrinsic chirality plays an important role in enantiomeric discrimination. Beyond the optimum helix height, the discrimination factor decreases due to weaker coupling of the photonic and plasmonic structure.

## III. CONCLUSIONS

We have thus reported an intrinsically chiral hybrid photonic-plasmonic metastructure exhibiting an exceptional point and demonstrated that the singularity enables giant enantiomeric discrimination. Near the EP singularity, small chiral perturbations produce strongly enhanced mode splittings that depend on the handedness of the perturbation, yielding more than an order-of-magnitude difference between opposite enantiomers and exploiting both frequency and loss-rate splittings. Beyond chiral sensing, our results establish intrinsic chirality as a new degree of freedom for controlling the response of non-Hermitian systems and open opportunities for exploring the interplay between chirality and non-Hermitian singularities in linear and non-linear regimes.